\documentclass{emulateapj} 
\usepackage{times}
\newcommand{\beq}{\begin{equation}}
\newcommand{\eeq}{\end{equation}}

\begin{document}

\shorttitle{\textsc  ACCRETION DISK CORONAE AND LARGE SCALE MAGNETIC FIELDS}
\shortauthors{\textsc BLACKMAN \& PESSAH}

\title{Coronae as Consequence of Large Scale Magnetic Fields in
  Turbulent Accretion Disks}
\author{Eric G. Blackman$^{1}$ and Martin E. Pessah$^{2}$}
\affil{1. Dept. of Physics and Astronomy, University of Rochester,
  Rochester NY 14627, USA; blackman@pas.rochester.edu \\ 2. Institute
  for Advanced Study, Princeton, NJ, 08540, USA; mpessah@ias.edu}

\begin{abstract}
Non-thermal X-ray emission in compact accretion engines can be interpreted to result from magnetic dissipation in an optically thin magnetized corona above an optically thick accretion disk. If coronal magnetic field originates in the disk and the disk is turbulent, then only magnetic structures large enough for their turbulent  shredding time to exceed their buoyant rise time survive the journey to the corona. We use this concept and a physical model to constrain the minimum fraction of magnetic energy above the critical scale for buoyancy as a function of the observed coronal to bolometric emission. Our results suggest that a significant fraction of the magnetic energy in accretion disks resides in large scale fields, which in turn provides circumstantial evidence for significant non-local transport phenomena and the need for large scale magnetic field generation. For the example of Seyfert AGN, for which $\sim$ 30 per cent of the bolometric flux is in the X-ray band, we find that more than 20 per cent of the magnetic energy must be of large enough scale to rise and dissipate in the corona.
\end{abstract}


\section{Introduction}
\label{sec:introduction}

Accretion disks are widely appreciated to be a powerful source of
emission from gas or plasma orbiting central stars or compact objects
\citep[see, e.g.,][]{FKR02}.  In order to explain the rapid
variability and short lifetimes of accreting systems without
unphysical mass densities, some enhanced angular momentum transport
beyond that which can be supplied by the microphysical transport
coefficients is typically required \citep{SS73}.  Many accreting
sources show jets, outflows, and active coronae highlighting that disk
dynamics and energy release involves some combination of local and
large scale transport.  Understanding the relative balance between the
two is of fundamental importance.

In some disk models, the primary angular momentum transport and
dissipation takes place above the disk (Lynden-Bell 1969; Field and
Rogers 1993) and the turbulence within the disk itself plays a
secondary role.  However, the magnetorotational instability (MRI) has
emerged as a likely source of turbulence within accretion disks, and a
leading candidate to contribute local turbulent angular momentum
transport in sufficiently ionized disks \citep{Velikhov59,
  Chandrasekhar60, BH91,BH98}.  Three-dimensional numerical
simulations \citep{HGB95, HGB96, Brandenburgetal95, Stoneetal96} have
revealed that the nonlinear evolution of systems unstable to the MRI
leads to sustained MHD turbulence and outward angular momentum
transport.

Understanding the saturation of the MRI is currently a topic of active
research \citep[see][and references therein]{PG09}. Local shearing box
simulations have not converged to practical angular momentum transport
coefficients, which are found to depend strongly on the simulation box
size and the initial seed magnetic field \citep[see, e.g.][]{HGB95,
  Sanoetal04, PCP07, Bodoetal08}, as well as the magnetic Prandtl
number \citep{FSH00, SI01, FPLH07, LL07}.  An important frontier in
this regard is to understand the development and role of large scale
magnetic field structures in the saturation of MRI-driven turbulence
\citep{MS00, Brandenburgetal95} and, more generally, in the process of
angular momentum transport. A realistic disk likely involves coupled
internal and coronal dynamics \citep[see, e.g.,][]{KB04}.

Observationally, the relevance of large scale magnetic fields in
accretion disks is strongly motivated by the interpretation of X-ray
flux in Seyferts which has been best interpreted as coronal emission.
The flux from 1-500 keV ranges from 10 to 50 per cent of the total
flux (Mushotzky et al. 1993).  Galactic black hole X-ray sources show
both thermal and non-thermal (power law) spectral components, with the
ratio of non-thermal to total luminosity ranging between 20 and 40 per
cent (Nowak 1995).  The leading paradigm for X-ray emission in these
accreting systems involves an optically thin, hot corona powered by
magnetic field dissipation (e.g. Haardt \& Maraschi 1993: Field \&
Rogers 1993).  If the corona results from magnetic structures
dissipating above the disk midplane that were originally produced
within the turbulent disk (e.g., via magnetic instabilities, such as
the MRI), then these structures must be of large enough scale to
survive the buoyant rise without being prematurely shredded by disk
turbulence.

If all coronal and jet emission results from fields initially produced
within a turbulent disk, then the fraction of coronal to bolometric
luminosity is directly related to the fraction of magnetic energy
associated with buoyant fields of large enough scale to survive the
vertical trip without being turbulently shredded.  In this
\emph{Letter} we employ this concept to develop a model  relating
the observed ratio of coronal to bolometric emission to the fraction of
magnetic energy produced in the disk that is of large enough scale to
buoyantly rise to the corona.  By comparing the implications of our
model with observations we infer that a significant fraction of
magnetic energy in accretion disks must reside in large scale fields.

\section{Why Coronae Require Large Scale Fields}

In order for magnetic fields to power coronae and jets, the buoyancy
time, $t_{\rm b}$, associated with a magnetic structure rising through
the disk must be smaller than the time associated with its turbulent
diffusion, $t_{\rm d}$.  These timescales are estimated as $t_{\rm b}
\equiv H/{U}_{\rm b}$, where $H$ is the disk half thickness and
${U}_{\rm b}$ is the characteristic buoyancy speed, and $t_{\rm d}
\equiv l^2/\nu_{\rm t}$ , where $\nu_{\rm t}\sim v l_{\rm t}$ is the
turbulent magnetic diffusion coefficient, $v$ is the dominant
turbulent speed, and $l_{\rm t}$ is the characteristic scale of a
typical (anisotropic) turbulent cell, i.e., $l_{\rm t} \sim \langle
l_xl_yl_z \rangle^{1/3}$.

The escape condition, $t_{\rm d} < t_{\rm b}$, sets a lower bound on
the scale $l$ of magnetic structures that can survive shredding and
reach the coronae (Blackman \& Tan 2003), namely
\begin{equation}
  l^2 >  l_{\rm c}^2 \equiv l_{\rm t} H \frac{v} {U_{\rm b}} \,.
\label{eq:l_gr_lc}
\end{equation}
The critical scale applies to the {\it smallest} dimension of a given
magnetic structure, i.e., $l_{\rm c} \le \min\{l_x,l_y,l_z\}$, since
this determines the shortest diffusion time. For a thin magnetic loop
$l$ would be the loop thickness not the distance between foot-points.
In the $\alpha$-viscosity disk framework \citep{SS73}, $\nu_{\rm t}
\equiv v l_{\rm t} \equiv \alpha c_{\rm s} H \sim v^2 /\Omega$, where
$c_{\rm s}$ is the sound speed and $\Omega$ is the local angular
frequency.  Using $c_{\rm s} \sim \Omega H$, this implies
$v=\alpha^{1/2}c_{\rm s}$ and thus $l_{\rm t} \sim \alpha^{1/2}H$.
Plugging these into Equation~(\ref{eq:l_gr_lc}) implies
\begin{equation}
  l_{\rm c}^2 \equiv  \alpha H^2 \frac{c_{\rm s}}{U_{\rm b}(l_{\rm c})} \,. 
\label{eq:lc}
\end{equation}
In a fully developed MHD/MRI turbulent flow the velocity of energy
containing turbulent motions is approximately (if not slightly less
than) the RMS Alfv\'en speed, i.e., $v \sim v_{\rm A}$.  Therefore if
$U_{\rm b}\le v_{\rm A}\sim v \sim \alpha^{1/2} c_{\rm s} $, then $l>
l_{\rm c} > \alpha^{1/4} H = l_{\rm t}/\alpha^{1/4}$ , which is larger
than $l_{\rm t}$ for $\alpha < 1$.  If instead we use $U_{\rm b}(l)\le
c_{\rm s}$, the analogous procedure gives $l> l_{\rm c} > \alpha^{1/2}
H = l_{\rm t}$.  Either way, the buoyant magnetic structures that
survive turbulent shredding in the disk must have $l>l_{\rm c}\ge
l_{\rm t}$.

That $l_{\rm c}$ equals or exceeds the characteristic turbulent scale
implies that an accretion engine with a significant fraction of power
emanating from coronae requires a significant fraction of magnetic
energy to be organized in magnetic structures of large scale with
$l>l_{\rm c}$.  In the sections that follow, we quantify this fraction
by determining $l_{\rm c}$ and $U_{\rm b}(l)$ and connect them with
the observed ratio of coronal to bolometric emission flux.

\section{ Coronal Emission Fraction}

We develop a model in which the observed fraction of coronal to
bolometric disk luminosity is determined by the rate of large scale
magnetic energy rising to the corona.  We assume that buoyant
structures fill a volume fraction $f_{\rm v}$ in the disk such that
the average disk mass density is given by
\begin{equation}
\rho \equiv \rho_o -f_{\rm v} (\rho_o-\rho_i) \,,
\label{eq:rho}
\end{equation}
where $\rho_o$ and $\rho_i$ are the mass densities
external to and internal to buoyant structures, respectively. The
average magnetic energy density is then
\begin{equation}
{B^2\over 8\pi} \equiv {1\over 8\pi}(B_o^2+f_{\rm v} B_{\rm ls}^2)\,,
\label{eq:B_sq}
\end{equation}
where ${B_{\rm ls}^2} \equiv {B_i^2 -B_o^2}$ corresponds to the
difference of internal and external magnetic fields squared.  We
suppose that although the entire magnetic energy of the structures
$B_i^2$ contributes to their initial buoyancy, only $B_{\rm ls}^2$
survives to the corona; the smaller scale fields are "bled" away
during the buoyant rise.  We define the fraction of the magnetic
energy density in scales larger than the critical scale for surviving
the buoyant rise to be
\begin{equation} 
  f_{\rm s} \equiv {f_{\rm v} B_{\rm ls}^2 \over B^2}
  ={f_{\rm v} B_{\rm ls}^2 \over f_{\rm v} B_{\rm ls}^2 + B_o^2}  \,.
\label{eq:fs}
\end{equation} 
The factor $f_{\rm v}$ arises in the numerator because $B_{\rm ls}^2$
is contained only in the volume of the buoyant structures.  The
quantity $f_{\rm s}$ represents the fraction of magnetic energy with
scales larger than $l_{\rm c}$, which can be written more generally in
terms of integrated magnetic spectra as
\begin{equation} 
  f_{\rm s}=
  \frac{\int^{k_{\rm c}}_{k_{\rm min}}   E_{\rm M}(k) dk}
  {\int^{k_{\rm max}}_{k_{\rm min}} E_{\rm M}(k) dk} \,, 
\end{equation} 
where the limiting wavenumbers are $k_{\rm c} = 2\pi/l_{\rm c}$,
$k_{\rm min} = 2\pi/H$, and $k_{\rm max} = 2\pi/l_{\rm diss}$ and
$l_{\rm diss}$ is the dissipation scale.

Motivated by the physical picture described above, we break up the
 accretion energy per unit area dissipated at a given radius
 into the sum of the dissipation associated with small scale field within the disk
\begin{eqnarray}
  D_{\rm d}  \equiv {Q\Sigma \Omega^2  \nu_{\rm t}}(1-f_{\rm s})  = 
  2Q  \alpha \beta c_{\rm s} \epsilon_{\rm mag} (1-f_{\rm s}) \,, 
\label{eq:D_d}
\end{eqnarray}
and the dissipation of large scale field in the corona
\begin{equation}
  D_{\rm c}  \equiv  f_{\rm s} U_{\rm b}  \epsilon_{\rm mag} \,.
\label{eq:D_c}
\end{equation}
Here, $\Sigma$ is the surface density, $Q\equiv (d\ln\Omega/ dR)^2/2$
and $\beta \equiv \rho c_{\rm s}^2/2 \epsilon_{\rm mag}$, where
$\epsilon_{\rm mag} \equiv B^2/8\pi$ is the total magnetic energy
density. The first term on the right of Equation~(\ref{eq:D_d})
resembles that which would follow from standard disk theory
\citep{FKR02} but with the extra factor of $1-f_{\rm s}$.
   
Defining the ratio of coronal to total dissipation as
\begin{equation}
  q\equiv \frac{D_{\rm c}}{D_{\rm d}+D_{\rm c}} \,,
  \label{q}
\end{equation}
the ratio of coronal to disk dissipation becomes
\begin{equation}
  \frac{D_{\rm c}}{D_{\rm d}} = {q\over 1-q}=  
  \frac{ f_{\rm s}}{2Q \alpha \beta
    (1-f_{\rm s} )}  \frac{U_{\rm b}}{c_{\rm s}} \,,
\label{eq:Dc_to_Dd}
\end{equation}
and depends cleanly on the ratio $U_{\rm b}/c_{\rm s}$.  Since both
$D_{\rm c}$ and $D_{\rm d}$ are expected to be dominated by their
contributions near the inner radius, we do not address the radial
dependence of these quantities in detail here and interpret $q$ as an
estimate of total coronal to bolometric emission.

\begin{figure}[t]
  \includegraphics[width=\columnwidth,trim=0 0 0 0]{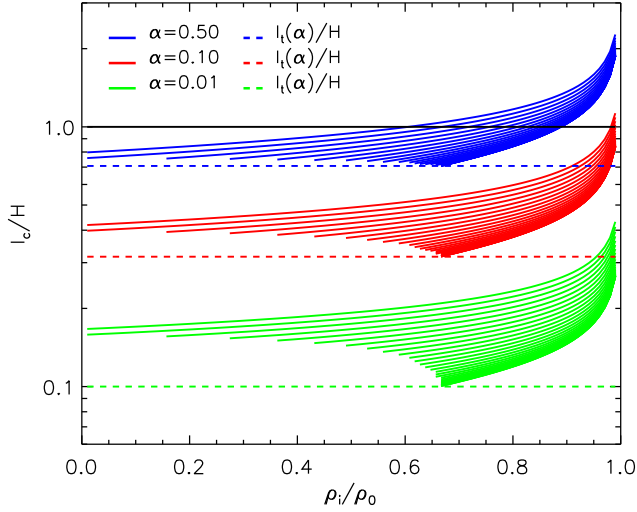}
  \caption{Ratio of the critical scale for buoyancy to the disk scale
    height, $l_{\rm c}/H$, that satisfies Eqs.~(\ref{eq:lc}) and
    (\ref{eq:Ub_to_cs}) simultaneously, as a function of the density
    ratio $\rho_i/\rho_o$.  The three sets of solid curves correspond
    to $\alpha = \{0.5, 0.1, 0.01\}$ from top to bottom.  Larger
    values of $\alpha$ correspond to more efficient shredding and thus
    require larger scales $l_{\rm c}$ to survive the buoyant rise.
    For a given $\alpha$, the solutions correspond to a range of drag
    coefficients $10^{-3} \le C_{\rm dr}\le 2$, logarithmically
    spaced.  The bottommost curves in each set correspond to the
    lowest drag; larger ${C_{\rm dr}}$ requires a larger $l_{\rm c}$
    to survive buoyant rise.  The dashed lines show the turbulent
    scale $l_{\rm t} =\alpha^{1/2}H$. That each set of curves lies
    above the line associated with $l_{\rm t}(\alpha)$ shows the
    importance of large scale fields.  A corona cannot form for region
    $l_{\rm c}>H$.}
  \label{fig:lc_to_H}
\end{figure}
\medskip

\section{Constraining the Buoyancy Speed }

To constrain the ratio $U_{\rm b}/c_{\rm s}$, we consider forces
on a magnetic structure in pressure balance with its exterior, i.e., 
\begin{equation}
  \frac{B_i^2}{8\pi} + n_i k_{\rm B} T_i   =  
  \frac{B_o^2}{8\pi} +  n_o k_{\rm B} T_o\,,
\end{equation}
where $n$ and $T$ are the corresponding number densities and
temperatures, and $k_{\rm B}$ is the Boltzmann constant.  Assuming
that $T_i=T_o=T$, we obtain
\begin{equation}
 \rho_o-\rho_i  = \frac{B_{\rm ls}^2}{8\pi c_{\rm s}^2} \,. 
\label{eq:rho_difference}
\end{equation} 

We take the force density acting on the magnetic structure to be
$F_{\rm b}-F_{\rm dr}$, where $F_{\rm b}$ is the upward gravitational
buoyancy force density and $F_{\rm dr}$ is the drag force density.
The former can be estimated according to
\begin{equation}
  F_{\rm b} \equiv g {H}  \frac{B_{\rm ls}^2}{8\pi c_{\rm s}^2}  =
  \frac{B_{\rm ls}^2}{8\pi H} \,, 
\label{eq:withdrag}
\end{equation}
where the last equality follows from hydrostatic equilibrium for an
isothermal gas ($\Gamma=1$), that is $k_{\rm B} T/m_{\rm p} = c_{\rm
  s}^2 = (GM/R)(H/R)$, with $g\equiv GM/R^2$.  We approximate $F_{\rm
  d}$ to be the high Reynolds number hydrodynamic drag associated to a
cylinder of length $L$ and diameter $l$, namely (see, e.g.
Moreno-Insertis 1986; Landau and Lifshitz 1987)
\begin{equation} 
  F_{\rm dr} \equiv {C_{\rm dr}\over 2}\rho_o U_{\rm b}^2
  \left[{lL\over \pi (l/2)^2 L}\right] =
  {2 C_{\rm dr}\over \pi l}\rho_o U_{\rm b}^2 \,,
\label{eq:Fd}
\end{equation} 
with drag coefficient $C_{\rm dr}$ of order unity.  The work done per
unit volume by the net force over a distance $H$ equals the kinetic
energy density of the rising structure, $H(F_{\rm dr}-F_{\rm b})\simeq
\rho_i U_{\rm b}^2/2$.  Combining this with
Equations~(\ref{eq:rho_difference})--(\ref{eq:Fd}) we obtain
\begin{equation} 
  {U_{\rm b}\over c_{\rm s}} = \left(1-{\rho_i\over \rho_o}\right)^{1/2} 
  \left({2C_{\rm dr}\over \pi}{H\over l} + {\rho_i\over 2\rho_o}\right)^{-1/2} \,.
\label{eq:Ub_to_cs}
\end{equation}
Therefore, for all buoyant structures $U_{\rm b}(l) \ge U(l_{\rm c})$
since $l\ge l_{\rm c}$.  Given that $l\le H$ for all structures that
fit in the disk, $U_{\rm b}(l) < c_{\rm s}$ for all $l$ would imply
$\rho_i/\rho_o > 2/3 - 4C_{\rm dr}/3\pi$.  We thus restrict ourselves
to this density ratio regime.  Note that our estimate for $U_{\rm
  b}/c_{\rm s}$ is itself an upper limit since we consider only a
hydrodynamic drag force restricting the buoyant rise, ignoring for
example, magnetic tension, c.f. \citet{ST02}.  Whether the dynamics
allows densities below the above upper limit remains an open question.
However, even if the structures were initially able to move faster
than $c_s$, we would expect shocks and the associated dissipation to
slow the motion via additional drag.

\begin{figure}[t]
  \includegraphics[width=\columnwidth,trim=0 0 0 0]{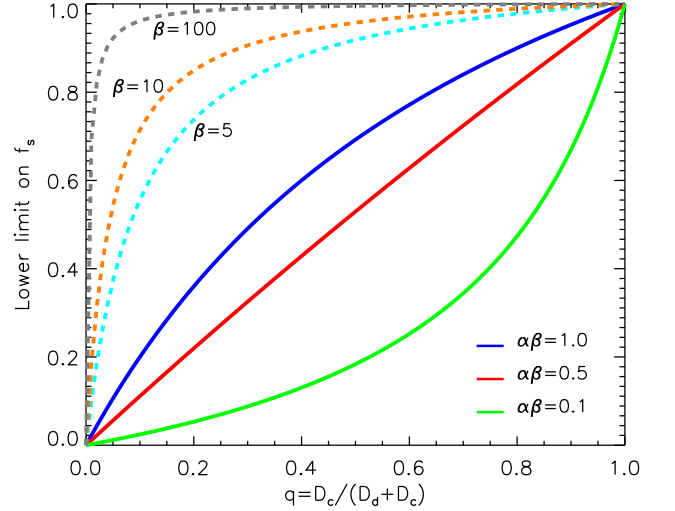}
  \caption{Lower limits on the fraction of magnetic energy residing at
    scales large enough for their buoyant rise time to be small
    compared to the corresponding turbulent shredding time, $f_{\rm
      s}$, as a function of the ratio of coronal to bolometric  dissipation,
    $q$. The solid curves correspond to $U_{\rm b}=c_{\rm s}$ in
    Eq.~(\ref{eq:Dc_to_Dd_2}) which gives the most stringent lower
    limit for $\alpha \beta=\{1.0, 0.5, 0.1\}$, from top to bottom.
    The dashed curves correspond to $U_{\rm b} = \alpha c_{\rm s}$, as
    suggested in Vishniac (1995).  In this case, the lower limit given
    by Eq.~(\ref{eq:Dc_to_Dd_2}) is independent of $\alpha$. The
    dashed lines correspond to $\beta = \{100, 10, 5\}$, from top to
    bottom ($\beta=1$ would coincide with the curve for
    $\alpha\beta=1$ for $U_{\rm b}=c_{\rm s}$). This last set of
    curves show that the same ratio of coronal to bolometric emission     requires a higher fraction of large scale fields as the buoyancy
    speed is reduced.}
  \label{fig:fs_vs_q}
\end{figure}

Vishniac (1995) estimated a buoyant rise time $U_{\rm b}\sim \alpha
c_{\rm s}$. This value is low compared to the upper limit $c_{\rm s}$
we considered. Had we used $U_{\rm b} \sim \alpha c_{\rm s}$, then the
value of $f_{\rm s}$ in Equation~(\ref{eq:fs_low_lim}) would
dramatically increase; highlighting the importance of large scale
field for coronal dynamics even more. Some characteristic values of
$f_{\rm s}$ resulting from this smaller buoyancy speed are shown as
dashed lines in Figure~\ref{fig:fs_vs_q}.

\section{Quantifying The  Importance of Large Scale Fields}

Setting $l=l_{\rm c}$ in Equation~(\ref{eq:Ub_to_cs}), and using
Equation~(\ref{eq:lc}), gives a $5^{\rm th}$ order equation for
$l_{\rm c}/H$ as a function of $C_{\rm dr}$, $\rho_i/\rho_o$, and
$\alpha$.  The physical solution to this equation is shown in
Figure~\ref{fig:lc_to_H} for a range in drag coefficients, $10^{-3}\le
C_{\rm dr} \le 2$, and for three values of $\alpha=\{0.5, 0.1, 0.01\}$
as a function of the density ratio $\rho_{\rm i}/\rho_o$.  The
characteristic turbulent scale $l_{\rm t} \equiv \alpha^{1/2}H$ is
also shown as horizontal lines for each value of $\alpha$. Since
$l_{\rm c}$ is the minimum scale for buoyant rise, the fact that
$l_{\rm c}> l_{\rm t}$ highlights the importance of large scale
fields. For the regime on the plot where $l_{\rm c}>H$, a buoyant
corona cannot arise from fields produced internally to the disk.

We can set two constraints on the minimum fraction $f_{\rm s}$ of
magnetic energy that the disk must produce in fields with scales
$l>l_{\rm c}$ for a given coronal to bolometric emission fraction $q$. The
more stringent bound relies on the fact that $U_{\rm b}(l) < U_{\rm
  b}(H)$ for $l<H$; a less severe limit is obtained requiring $U_{\rm
  b}(H)<c_{\rm s}$, which is satisfied as long as the density ratio
$\rho_i/\rho_o$ does not fall below the lower limit discussed above.
Applying these conditions to Equation~(\ref{eq:Dc_to_Dd}) we obtain
\begin{equation}
  {D_{\rm c}\over D_{\rm d}} 
  ={q\over 1-q}
  \le  {f_{\rm s} U_{\rm  b}(H)/c_{\rm s} \over 2Q \alpha \beta (1-f_{\rm s} )}
  \le  {f_{\rm s} \over 2Q \alpha \beta (1-f_{\rm s} )} \,.
 \label{eq:Dc_to_Dd_2}
\end{equation}
Then, for an observationally inferred value of coronal to bolometric
flux, we can obtain a lower limit on the fraction of magnetic energy
residing in large scale fields, i.e.,
\begin{equation}
  f_{\rm s} \ge \frac{\alpha \beta}{\alpha\beta + 4(1/q-1)/9} \,,
\label{eq:fs_low_lim}
\end{equation}
where we have used that $Q\sim 9/8$ for a Keplerian disks.

The lower limit for $f_{\rm s}$ depends on the dimensionless
parameters characterizing the angular momentum transport efficiency,
$\alpha$, and magnetic pressure support, $\beta$, \emph{only} though
the product $\alpha\beta$.  This result is very encouraging because,
despite the fact that both quantities vary over several orders of
magnitude across simulations carried out in domains with various sizes
and with different field strengths and geometries \citep[see][and
references therein]{PCP06a}, their product remains nearly constant
\footnote{The constancy of $\alpha\beta$ is consistent with the
  relations below Equation~(\ref{eq:lc}); for a disk with MRI growth
  time $\sim \Omega^{-1}$ we expect $\nu_{\rm t}=\alpha c_{\rm s}
  H\sim v^2/\Omega \sim v_{\rm A}^2/\Omega$.  Using $c_{\rm s} \sim
  \Omega H$ then gives $\alpha \propto \beta^{-1}$ with a
  proportionality constant dependent upon anisotropy and the
  polytropic index \citep{BPV08}.} with $\alpha \beta \simeq 0.5$
\citep[see][and references therein]{BPV08}.

The upper limits for the fraction of magnetic energy associated with
large scale field structures are shown in Figure~\ref{fig:fs_vs_q} for
three values of the product $\alpha\beta = \{1.0,0.5,0.1\}$.  The
minimum constraints on $f_{\rm s}$ illustrated show that, if the
observed non-thermal emission is interpreted as coronal emission due
to magnetic dissipation from buoyant fields that were produced within
a turbulent disk, then a significant fraction of the energy budget of
the magnetic field build in the disk must be produced in fields of
scale $l>l_{\rm c}$. Since Figure~\ref{fig:lc_to_H} shows that in
general $l>l_{\rm t}$, together these figures highlight the importance
of large scale magnetic fields in powering coronae.

Finally we note that MHD jets models typically invoke global scale
fields, c.f. \citet{f07}.  If these fields arise from the opening of
coronal fields (as in the sun; Wang \& Sheeley 2003: Blackman \& Tan 2004) then the
mechanical luminosities of jets would represent an additional
contribution to that which results from the buoyant rise of magnetic
fields.  Specifically, $D_c$ in Equation (\ref{q}) would be repalced
by $D_c +D_j$ where the latter is the jet power.  This would further
increase our lower limits on $f_s$.

\section{Relation to Previous Work}

In our physical picture, the coronal emission fraction $q$ depends on
the fraction of magnetic energy $f_{\rm s}$ produced in scales larger
than the critical scale $l_{\rm c}$.  We incorporate the density
contrast between buoyant structures and the ambient medium required
for coronal feeding.  Also, our lower limit on $f_{\rm s}$ employs
$c_s$ as an upper limit for the buoyant rise time.  We use the
$\alpha$-viscosity prescription only for dissipation inside the disk;
the coronal dissipation is modeled as a distinct contribution (see
Eqs.~[\ref{eq:D_d}] and [\ref{eq:D_c}]).  These features differ from
those of Merloni \& Fabian (2002) and the ones summarized in Wang et
al.~(2004). In those treatments, there is no explicit distinction
between the density inside and outside the buoyant structures or the
role of large and small scale magnetic fields.  Furthermore, the
coronal emission fraction is considered to be a subset of the total
dissipation, modeled entirely with the $\alpha$-viscosity
prescription. Also, the buoyant rise time is taken to be the Alfv\'en
speed; less than our upper limit value of $c_{\rm s}$.

In the present work we implicitly consider systems with low enough
accretion rates such that radiation pressure is unimportant.  A
subtlety associated with radiation pressure is that the thermal
photosphere can be significantly higher than the scale at which the
magnetic pressure dominates the thermal pressure (e.g.  Hirose et
al.~2009).  Thus the non-thermal coronal emission arises at larger
scale heights compared to when radiation pressure is ignorable.  The
reduction of the coronal emission fraction for such large accretion
rate systems is observed in Wang et al. (2004).  For a fixed magnetic
spectrum, this would also be expected in our paradigm, because $H$ in
Equation~(\ref{eq:lc}) is the scale that we consider a buoyant
structure must rise to contribute to coronal emission.  If the buoyant
structure has to move higher, then $l_{\rm c}$ would be larger and
less of the magnetic energy would survive the buoyant rise.  In
Merloni \& Fabian (2002) the reduction in coronal emission for large
radiation pressure occurs because their coronal emission fraction
depends inversely on the total disk pressure.

\section{Conclusions}

Starting with the assumption that coronal luminosity from a turbulent
accretion disk results from buoyant magnetic structures that survive
turbulent shredding for at least one vertical density scale height, we
derived lower limits on: ({\it i}) the scale of such magnetic
structures and ({\it ii}) the fraction of magnetic energy that needs
to be produced above this scale within the disk to account for
observed values of coronal to bolometric luminosity. In our minimalist
model, we considered the buoyant structures to be in pressure
equilibrium with the ambient medium but to have an additional magnetic
energy contribution from scales above the critical scale $l_{\rm c}$,
and a lower density.

We find that typical ratios of coronal to bolometric luminosity
observed in AGN require the critical scale for buoyancy to robustly
exceed the characteristic scale set by turbulent motions and that
double digit percentages of magnetic energy should reside in fields
above this scale.  This is consistent with recent work highlighting
the importance of in situ large scale dynamos in feeding coronae
(Blackman 2007; Vishniac 2009).  Our results complement growing
motivation to consider larger domains in stratified MRI simulations
and motivate analysis of the magnetic energy spectra produced therein.
The results also resonate with models of accretion disks in which
buoyancy and coronal dissipation play a primary role for transport
(Lynden-Bell 1969; Field \& Rogers 1993).

\acknowledgments{We thank J. Goodman, A. Hubbard, V. Pariev, and D.
  Uzdensky for related discussions.  EGB acknowledges NSF grants
  AST-0406799, AST-0406823, and NASA grant ATP04-0000-0016 and the LLE
  at UR.  MEP gratefully acknowledges support from the Institute for
  Advanced Study.}

\end{document}